\documentstyle[aps]{revtex}

\def\half{\textstyle{1 \over 2}}
\def\dyv{\mbox{div}}
\def\Re{\mbox{Re}}
\def\Im{\mbox{Im}}
\def\jev{\mbox{{\boldmath $\jmath$}}}
\def\jav{\langle\mbox{{\boldmath $\jmath$}}\rangle}
\def\cdott{\mathord{\cdot}}
\def\taub{{\mbox{\boldmath ${\tau}$}}}
\def\stress{\mathord{\taub{\mkern-10.8mu}\taub}}
\def\uvec{\mbox{{\boldmath $u$}}}
\def\vvec{\mbox{{\boldmath $v$}}}
\def\nvec{\hat{\mbox{{\boldmath $n$}}}}
\def\rvec{\mbox{{\boldmath $r$}}}
\def\tvec{\hat{\mbox{{\boldmath $t$}}}}

\def\Fvec{\mbox{{\boldmath $F$}}}
\begin{document}
\draft
\title{Anomalous acoustic reflection on a sliding interface or a shear band}

\author{C. Caroli, B. Velick\'y\cite{Charles} }
\address{Groupe de Physique des Solides\cite{CNRS}, 2 place Jussieu,
75251 Paris cedex 05, France.}
\date{\today}
\maketitle

\begin{abstract}
We study the reflection of an acoustic plane wave from a steadily 
sliding planar interface with velocity strengthening friction or a 
shear band in a confined granular medium. The corresponding acoustic 
impedance is utterly different from that of the static interface. In 
particular, the system being open, the energy of an in-plane polarized 
wave is no longer conserved, the work of the external pulling force 
being partitioned between frictional dissipation and gain (of either 
sign) of coherent acoustic energy. Large values of the friction 
coefficient favor energy gain, while velocity strengthening tends to 
suppress it. An interface with infinite elastic contrast (one rigid 
medium) and V-independent (Coulomb) friction exhibits spontaneous 
acoustic emission, as already shown by M. Nosonovsky and G.G. Adams 
(Int. J. Ing. Sci., {\bf 39}, 1257 (2001)). But this pathology is 
cured by any finite elastic contrast, or by a moderately large 
V-strengthening of friction.

We show that (i) positive gain should be observable for rough-on-flat 
multicontact interfaces (ii) a sliding shear band in a granular medium 
should give rise to sizeable reflection, which opens a promising 
possibility for the detection of shear localization.

\end{abstract}  


\section{Introduction}
The question of the origin and nature of shear localization
in disordered systems, such as soft glasses or confined granular
media, which are jammed at equilibrium, but flow when sheared beyond
a threshold stress, is a long standing one. Due in particular to recent
progress in
theoretical descriptions, it is presently the subject of renewed
interest. Hence the need for identifying appropriate, non invasive methods of
experimental investigation which could complement the optical and NMR
imaging ones, recently put to use by Pignon et al \cite{Piau}, and by
Coussot et al \cite{Coussot}. We intend to show in this article
that the propagation of sound
pulses appears as a promising possibility. Namely, we will show that
the presence of a shear band of thickness small compared with the
acoustic wavelength should give rise to strong anomalous reflection
of a transverse acoustic signal
for well defined incidence ranges. Such a method could therefore provide a
relatively handy fingerprint of shear localization in confined granular media.

An extreme case of localized shear flow is that of frictional
sliding of the interface between two cohesive macroscopic solids. In
such a configuration, the very structure of the system prelocalizes
shear to the nanometer-thick layer which forms the molecular adhesive
contact(s). The role of the above-mentioned threshold stress is
played here by the so-called static friction force. Below this
threshold, the interface is elastically pinned (jammed), and responds
elastically to a shearing force. So, the corresponding mechanical
boundary condition is simply that the displacement fields in the two
media must be fully continuous across the interface. But, beyond this
shear level, sliding sets in and, along the direction of relative
motion, the boundary condition is now provided by the dynamic
friction law, which states that the shear and normal stresses are
proportional. Obviously, such a discontinuity of the boundary
conditions must result in a discontinuous change of the acoustic
reflection and transmission coefficients when the system is set into
sliding.

This was pointed out already long ago by Chez et al \cite{Comninou}, who
studied
the reflection of a sound wave propagating in a plane orthogonal to
the sliding direction, and polarized in the plane of incidence on an
interface with the Coulomb frictional behavior (constant friction
coefficient). However, due to the choice of this particular geometry, they
overlooked an interesting effect. Indeed, the sliding system
is an open one : energy is being pumped
in from an external source -- the external driving machine which
imposes the sliding velocity. So, as
soon as the acoustic displacement field has a non-zero component, in
the interfacial plane, along the direction of the sliding motion,
additional mechanical work (of a priori either sign) is extracted from
the external source, and interfacial acoustic scattering is no longer
energy-conserving. This opens in principle the possibility of acoustic
gain at reflection, i.e. conversion of incoherent into coherent
mechanical energy -- quite an exciting prospect indeed!

Now, from the point of view of the propagation of an acoustic signal of
wavelength $\lambda$, a shear band of thickness $d<<\lambda$
in a confined granular medium appears as the equivalent of
a frictional interface between two identical solids. Indeed, the band
can then be considered as a surface of mechanical
discontinuity between the non sliding adjacent regions, which behave as
(disordered) elastic solids. Experimental studies by rock
mechanicians \cite{Tullis} \cite{Marone} of systems constituted
of two bulk rock pieces separated
by an interposed layer of granular material (called "gouge") have established
that in such systems (i) sliding occurs in a narrow band within the
gouge (ii) the dynamics is ruled by a standard solid friction law,
the associated friction coefficients having a magnitude
comparable with those for solids in direct contact.

Reflection and transmission of waves with a polarization component
along the sliding direction of an interface with constant
friction coefficient between dissimilar media
have been recently studied by Nosonovsky and Adams \cite{Noso1},
though in a different perspective. Namely, they focussed primarily on
the possible generation of slip pulses - a dynamical feature that seems
to be specific of pure Coulombic friction. In the course of this
article we will rederive some of their results, which will be extended
to the more realistic case of velocity-dependent friction and to the
shear band problem..

This article is organized as follows.

In Section 2, we first
write down the equations for the most general case of a
monochromatic acoustic wave incident upon a planar frictional
interface between two semi-infinite solids with different elastic
moduli. We then
specialize in Section 3 to the case where the elastic contrast between
the two media
is very large (e.g. a gel sliding on top of a glass plate). We show
that, if the stiffer medium is assumed strictly non deformable and
friction taken to be Coulomb-like, the reflection coefficient of a
wave with the sliding direction and polarization in the plane of incidence
is highly pathological : not only is a huge gain at reflection
possible for some particular incidence range, but spontaneous
acoustic emission from the surface is
predicted - a result already obtained recently by Nosonovsky and Adams
\cite{Noso2}.
These singularities are to be related with those already
found by Adams \cite{Adams}, and Ranjith and Rice \cite{Ranjith},
in their studies of interfacial
waves in the same system. They result, as is well known in mechanics,
from the specific singular character which the Coulomb model,
which takes the friction
coefficient to be a mere constant, shares with the Hill model of
plasticity.
Indeed, we show that (i) a very small finite relative compliance
of the stiffer medium is sufficient to destroy the acoustic
emission singularity (ii) improving upon the Coulomb description
by taking into account a velocity-strengthening dependence
of the dynamic friction coefficient of the order of what is measured
for real systems also cures this singularity. Moreover,
the possibility of energy gain at reflection is found to be strongly
dependent upon the strength of the velocity dependence, and hence on
the type of system : while it should be negligible for a gel-on-glass
system, it might be observable with a rough-on-flat multicontact
interface in well defined incidence ranges.

Section 4 is devoted to the symmetric case of two mechanically
identical solids relevant to the shear band problem. In
this situation,  and when the velocity dependence of the
friction coefficient is taken into account, we predict that the
sliding
shear band should give rise to a clear acoustic signature - namely a
reflection coefficient with magnitude of order typically of order
unity for incident signals in well defined ranges of incidence angles.

\section {General formulation}

Consider two elastically isotropic semi-infinite media (M; M') with
shear moduli and Poisson ratios ($\mu,\nu~;~\mu',\nu'$) and densities
($\rho, \rho'$),
occupying respectively (Figure \ref{fig:1})
 the upper ($x_{2}>0$) and lower ($x_{2}<0$) half spaces, and in
continuous contact with each other along the $x_{2}=0$ plane. Medium
(M')
is assumed to be in stationary sliding motion with respect to (M) at
velocity $v$ along $x_{1}$ and towards $x_{1}<0$. In this reference
state, the (homogeneous) normal and shear stresses $\tau^{*}_{22}$,
$\tau^{*}_{12}$, are related by the dynamic friction law :

\begin{equation}
\label{eq:friction law}
\tau^{*}_{12} =  -\,f(v)\tau^{*}_{22}
\end {equation}
with $f(v)$ the friction coefficient.

An emitter linked to (M) is sending from infinity towards the
interface a plane acoustic wave of frequency $\omega$, propagating in
the $x_{1}, x_{2}$ plane at incidence angle $\theta$ (Figure \ref{fig:1}) and
polarized in the plane of incidence. That is, the associated
displacement has a finite component along the sliding direction. In
order to fix ideas, and for the sake of simplicity, we restrict in
most of what follows the algebraic formulation to the case of a
transverse (shear) incident wave - the case of an incident
longitudinal (dilatational) signal follows straightforwardly. The
elastic displacement field ($u_{1}, u_{2}$) in (M) obeys the Lam{\'e}
equations :
\begin{equation}
\label{eq:Lame1}
\ddot{u_{1}} = c_{L}^{2}\,\frac{{\partial}^{2} u_{1}}{\partial
{x_{1}}^{2}} + (c_{L}^{2}-c_{T}^{2})\,
\frac{{\partial}^{2}u_{2}}{\partial x_{1}\partial x_{2}} +
 c_{T}^{2}\,\frac{{\partial}^{2} u_{1}}{\partial
{x_{2}}^{2}}
\end{equation}

\begin{equation}
\label{eq:Lame2}
\ddot{u_{2}} = c_{L}^{2}\,\frac{{\partial}^{2} u_{2}}{\partial
{x_{2}}^{2}} + (c_{L}^{2}-c_{T}^{2})\,
\frac{{\partial}^{2}u_{1}}{\partial x_{1}\partial x_{2}} +
 c_{T}^{2}\,\frac{{\partial}^{2} u_{2}}{\partial
{x_{1}}^{2}}
\end{equation}
$c_{L}, c_{T}$ are the longitudinal and transverse sound velocities
in (M), and

\begin{equation}
\label{eq:vitesse son}
c_{T}^{2} = \frac{\mu}{\rho}\,\,\,\,\,\,\,\,\,\,\,\,\,\,
\frac{c_{L}^{2}}{c_{T}^{2}} = \frac{2(1 - \nu)}{1 - 2\nu}
\end{equation}
The solution of equations (\ref{eq:Lame1}), (\ref{eq:Lame2})
reads :

\begin{equation}
\label{eq:u1}
u_{1} = \epsilon e^{i(kx_{1} - \omega t)}\left[e^{-iq_{T}x_{2}} + \alpha
e^{iq_{T}x_{2}} + \beta  e^{iq_{L}x_{2}}\right]
\end{equation}

\begin{equation}
\label{eq:u2}
u_{2} = \epsilon e^{i(kx_{1} - \omega
t)}\left[\frac{k}{q_{T}}e^{-iq_{T}x_{2}} - \alpha
\frac{k}{q_{T}} e^{iq_{T}x_{2}} + \beta \frac{q_{L}}{k}
e^{iq_{L}x_{2}}\right]
\end{equation}
where $\epsilon$ specifies the amplitude of the incident wave, $k$ is the
$x_{1}$ component of its wavevector, and

\begin{equation}
\label{eq:qLT}
q_{T,L}^{2} = \frac{\omega^{2}}{c_{T,L}^{2}} - k^{2}
\end{equation}
with $q_{T}$ real positive and $\Re\,q_{L} \geq 0$ and $\Im\,q_{L} \geq 0$.

In the lower medium (M'), which moves at velocity $(-v)$ in our
reference frame, analogously :

\begin{equation}
\label{eq:u'1}
u'_{1} = \epsilon e^{i(kx_{1} - \omega t)}\left[ \alpha'
e^{-iq'_{T}x_{2}} + \beta'  e^{-iq'_{L}x_{2}}\right]
\end{equation}

\begin{equation}
\label{eq:u'2}
u'_{2} = \epsilon e^{i(kx_{1} - \omega t)}\left[ \alpha' \frac{k}{q'_{T}}
e^{-iq'_{T}x_{2}} - \beta' \frac{q'_{L}}{k}
e^{-iq'_{L}x_{2}}\right]
\end{equation}
where
\begin{equation}
\label{eq:omega'}
\omega' = \omega + vk\,\,\,\,\,\,\,\,\,\,\,\,\,\,\,{ q'_{T,L}}^{2} =
\frac{{\omega'}^{2}}{{c'}_{T,L}^{2}} - k^{2}
\end{equation}

In order to determine the amplitude reflection ($\alpha, \beta$) and
transmission ($\alpha', \beta'$) coefficients into the transverse and
longitudinal channels we must now specify the four boundary
conditions (BC) to be satisfied along the {\it deformed} M/M' interface.

Since the acoustic stresses we consider are small, ${\cal
O}(\epsilon)$, we expect the contact to persist everywhere, so that

(i) normal displacements on both sides of the interface are equal :
$\uvec\cdot{\nvec}=\uvec'\cdot{\nvec}$
and mechanical equilibrium imposes continuity of normal and shear
stresses :

(ii) $\tau_{nn} = {\tau'}_{nn}$

(iii) $\tau_{nt} = {\tau'}_{nt}$

The sliding velocity perturbation $(v_{I} - v)$ being also ${\cal
O}(\epsilon)$, $v_{I}$ remains positive (sliding persists) everywhere,
and we assume that the dynamic friction condition now holds {\it
locally}, namely that, at each interfacial point :

(iv) $\tau_{nt} + f(v_{I}) \tau_{nn} = 0$

These boundary conditions, which have been taken for granted in existing
studies of interfacial waves and shear fractures, actually imply
important physical assumptions, which should be made explicit.

On the one hand, the (macroscopic) contact is supposed to be
continuous and homogeneous - hence the statement of homogeneous
stresses in the reference state. Now, it is well known that such is
not the prevalent case. Most interfaces, being formed between rough
solids, are actually constituted of a large, but rather sparse, set of
microcontacts \cite{Dieterich} \cite{Bo}, in between which the surfaces
are mechanically free. Then,
it is clear that conditions (i)-(iii) are valid only in a
coarse-grained sense, i.e. provided that the length scale $2\pi/k$ of
variation of the acoustic fields along $x_{1}$ is much larger than the
average distance $d$ between microcontacts, that is for $kd~<<~1$.

Intercontact distances lie commonly in the range of hundreds of
micrometers. Only in the case where one at least of the solids is extremely
compliant are macroscopic contacts truly continuous. This is for
example the case for elastomers or gels.

On the other hand, the mere existence of a finite static threshold
proves that frictional dissipation results from the triggering by the
applied shear of structural instabilities. It is now documented that
the corresponding  structural rearrangement events take place in the
nanometer-thick adhesive interfacial layers, and affect volumes of,
typically, nanometric scale \cite{Bo} \cite{Pauline2} - comparable with
the "shear transformation zones " invoked by Falk and Langer
\cite{Falk} to model the
plasticity of amorphous solids. Let us call $b$ the size of such a
zone. A friction law represents the result of a {\it statistical
average} over a large number of such dissipative events. So, it can
only make sense on a scale much larger than $b$, that is for $kb~<<~1$.

All this means that boundary conditions (i)-(iv) above must be understood
as valid only on scales larger than some {\it cut-off length} $L =
\max (b,d)$ on the order of, typically, (i) a fraction of micrometer
for conforming
solid contacts (ii) a fraction of millimeter for the, more common,
multicontact interfaces.

Finally, since $v_{I}$ now becomes a time-dependent quantity,
condition (iv) implies that we assume the friction relation to hold
instantaneously on the scale $\omega^{-1}$. It is known that this
does not apply to the
case where the steady state $f(v)$ is velocity weakening ($df/dv <0$).
Indeed, such behavior necessarily results from the action of some
underlying structural dynamics leading to aging when sticking and
rejuvenation upon sliding, such as that associated with the slow creep
growth of the real area of contact relevant to multicontact
interfaces\cite{RR} \cite{Heslot}. Then, the equations describing non steady
friction must
involve explicitly at least one more dynamical "state" variable, and
condition (iv) becomes insufficient. Note also that it is in this
regime that steady sliding may be unstable with respect to stick-slip.

So, we restrict ourselves in what follows to the
velocity-strengthening case $df/dv > 0$.
This can be expected to hold, for
rough-on-rough interfaces, only at sliding velocities in the
mm.sec$^{-1}$ range, large enough
for rejuvenation effects to be saturated \cite{Pauline1}. However,
it has been shown \cite{Bureau}
to prevail down to the $\mu$m.sec$^{-1}$ range when working with
rough-on-flat interfaces where contacts keep their identity when
sliding, which makes contact area saturation easily realizable.

The position of our deformed interface is given by

\begin{equation}
\label{eq: position interf}
x_{2I}\left(x_{1},t\right) =
u_{2}\left(x_{1}-u_{1}\left(x_{1},0,t\right)\right)
\end{equation}
We assume from now on that acoustic deformations are small enough for
us to work in the linearized approximation. Then, the normal and
tangent unit vectors $\nvec, \tvec$ are simply :

\begin{equation}
\label{eq:nt}
\nvec(x_{1},t) = \left(\begin{array}{c}-{\left(\frac{\partial
u_{2}}{\partial
x_{1}}\right)}_{x_{1},0,t}\\1\end{array}\right)\ \ \ \ \ \ \
\tvec(x_{1},t) = \left(\begin{array}{c}1\\{\left(\frac{\partial
u_{2}}{\partial
x_{1}}\right)}_{x_{1},0,t}\end{array}\right)
\end{equation}

Let the stress field in, say, M, be denoted $\tau^{*}_{ij} +
\delta\tau_{ij}$, ($i,j = 1,2$), with

\begin{equation}
\label{eq:contraintes}
\delta\tau_{ij} = \mu\left[\frac{\partial {u}_{i}}{\partial {x}_{j}} +
\frac{\partial {u}_{j}}{\partial {x}_{i}} + \frac{2\nu}{1-2\nu}
\delta_{ij} u_{kk}\right]
\end{equation}

Then, to first order, at the interface in M

\begin{equation}
\label{eq:taunn}
\tau_{nn} = n_{i}\tau_{ij}n_{j} = \tau^{*}_{22} + \delta\tau_{22} -
2\tau^{*}_{12}\left(\frac{\partial u_{2}}{\partial x_{1}}\right)
\end{equation}

\begin{equation}
\label{eq:taunt}
\tau_{nt} = \tau^{*}_{12} + \delta\tau_{12}
+\left(\tau^{*}_{22}-\tau^{*}_{11}\right)
\left(\frac{\partial u_{2}}{\partial x_{1}}\right)
\end{equation}
and the local sliding velocity

\begin{equation}
\label{eq:velocity}
v_{I} = v + (\dot{u}_{1}-\dot{u}'_{1})
\end{equation}

Conditions (i)-(iv) then become:

\begin{equation}
\label{eq:BC1}
{\left[u_{2}-u'_{2}\right]}_{x_{1},0,t} = 0
\end{equation}

\begin{equation}
\label{eq:BC2}
{\left[\delta\tau_{22} -\delta\tau'_{22}
-2\tau^{*}_{12}\left( \frac{\partial u_{2}}{\partial x_{1}}
- \frac{\partial u'_{2}}{\partial x_{1}}\right)\right]}_{x_{1},0,t} = 0
\end{equation}

\begin{equation}
\label{eq:BC3}
{\left[\delta\tau_{12} - \delta\tau'_{12} + \left(\tau^{*}_{22} -
\tau^{*}_{11}\right)
\left( \frac{\partial u_{2}}{\partial x_{1}}
- \frac{\partial u'_{2}}{\partial x_{1}}\right)\right]}_{x_{1},0,t} = 0
\end{equation}

\begin{equation}
\label{eq:BC4}
{\left[\delta \tau_{12}+f(v)\delta\tau_{22} -\tau^{*}
\frac{\partial u_{2}}{\partial x_{1}}+ f'(v) \tau^{*}_{22}
\left({\dot u}_{1} -{\dot u}'_{1}\right)
\right]}_{x_{1},0,t} = 0
\end{equation}

where $f'(v) = df/dv$, and we have set

\begin{equation}
\label{eq:taustar}
\tau^{*} = -\left(\tau^{*}_{22} -
\tau^{*}_{11} - 2f(v) \tau^{*}_{12}\right)
\end{equation}
$\tau^{*}_{22} < 0$ (compressive normal stress),
$\tau^{*}_{12} = - \tau^{*}_{22} > 0$ , and we can reasonably
assume sliding to occur under zero lateral stress, i.e.
 $\tau^{*}_{11} = 0$, so that, in general, $\tau^{*}> 0$.

 Note that the $\tau^{*}$-dependent terms in eqs (\ref{eq:BC1})-
(\ref{eq:BC4}), which account for the fact that the BC's must be
enforced along the {\it deformed} interface, have usually been
overlooked in previous works on interfacial waves. As will appear
below, they give rise to corrections on the order of
$(\tau^{*}/\mu)$. In the case of hard materials, externally applied
stresses are in practice always considerably smaller than elastic
moduli, and these corrections can safely be neglected. However, such
may not be the case when dealing with very compliant materials such as
gels or elastomers. For example, for the case of gel/glass interfaces
\cite{Ronsin}
sliding stress levels may be a sizeable fraction of $\mu$,
and the full expressions should be retained.

Then, using eqs (\ref{eq:u1}), (\ref{eq:u2}), (\ref{eq:u'1}),
(\ref{eq:u'2}) and (\ref{eq:contraintes}), the BC's yield the
following set of
four non homogeneous linear equations for $\alpha,\alpha', \beta,\beta'$ :

\begin{equation}
\label{eq:A1}
-\frac{k}{q_{T}}\alpha  +\frac{q_{L}}{k}\beta -
\frac{k}{q'_{T}}\alpha' +\frac{q'_{L}}{k}\beta' = -\frac{k}{q_{T}}
\end{equation}

\begin{equation}
\label{eq:A2}
-2\mu \alpha +\mu \frac{q_{T}^{2} -k^{2}}{k^{2}} \beta +2\mu'\alpha' -
\mu'\frac{{q'}_{T}^{2} -k^{2}}{k^{2}}\beta' = 2\mu
\end{equation}

\begin{equation}
\label{eq:A3}
\mu \frac{q_{T}^{2}-k^{2}}{q_{T}} \alpha + 2\mu q_{L} \beta +
\mu'\frac{{q'}_{T}^{2} -k^{2}}{q'_{T}} \alpha' + 2\mu'q'_{L} \beta' =
\mu \frac{q_{T}^{2}-k^{2}}{q_{T}}
\end{equation}

\begin{eqnarray}
\label{eq:A4}
\left[\frac{q_{T}^{2}-k^{2}}{q_{T}}- 2fk +\frac{\tau^{*}}{\mu}
\frac{k^{2}}{q_{T}}\right] \alpha + \left[ 2q_{L}
+f\frac{q_{T}^{2}-k^{2}}{k} - \frac{\tau^{*}}{\mu} q_{L}\right]
\beta  \hspace{80pt}\nonumber\\
\hspace{80pt}+ \frac{Af\omega}{c_{T}}\left(\alpha +\beta
-\alpha'-\beta'\right)=
\left[\frac{q_{T}^{2}-k^{2}}{q_{T}}+ 2fk +\frac{\tau^{*}}{\mu}
\frac{k^{2}}{q_{T}}\right] - \frac{Af\omega}{c_{T}}
\end {eqnarray}

where we have defined :

\begin{equation}
\label{eq:A5}
A = \frac{f'(v)c_{T}}{f(v)}\frac{\mid \tau^{*}_{22}\mid }{\mu}
\end {equation}
which measures the dimensionless strength of the
velocity-strengthening effect.

Equations (\ref{eq:A1})-(\ref{eq:A4}) above are relevant to the case 
of a transverse incident wave. We will also display some results for 
a longitudinal one. In this latter case, the only modification 
concerns the first terms in the r.h.s. of equations (\ref{eq:u1}), 
(\ref{eq:u2}) which give the expression of the total acoustic 
displacement field. This results in leaving the l.h.s.'s of the final 
equations (\ref{eq:A1})-(\ref{eq:A4}) unchanged, only the r.h.s.'s 
being modified. 

Since the general solution of our problem 
depends on a large number of parameters (four elastic moduli, two
densities, the angle of incidence, the friction coefficient and its
derivative) it will be more illuminating to focus on a few simple
cases, namely that of very large elastic contrast (quasi non
deformable M'), and the symmetric case of two identical materials,
relevant to the problem of shear band detection.

\section{Large elastic contrast case}

We consider here the case of maximum asymmetry, where M' is
considerably stiffer than M. In order to disentangle the respective
effects of the elastic contrast ratio $ R = \mu'/\mu $ and of the velocity
dependence of friction, we first treat the extra simple limit of a
non deformable M' and of pure Coulomb friction, then examine how the
results thus obtained are modified (i) when $R$ is large but finite
(ii) when $f'(v) \neq 0$.

\subsection{Non deformable M' and Coulomb friction\label{basic_case}}
The boundary conditions reduce to imposing that $u_{2} = 0$ and
$\tau_{nt}+f\tau_{nn} = 0$ along the $x_{2} = 0$ plane. Equations
(\ref{eq:A1})-(\ref{eq:A4}) become, in this $R^{-1} = f'= 0$ limit,
where $\alpha' = \beta' = 0$ :

\begin{equation}
\label{eq:rigCb1}
\frac{k}{q_{T}}\alpha - \frac{q_{L}}{k}\beta = \frac{k}{q_{T}}
\end{equation}

\begin{equation}
\label {eq:rigCb2}
\left[ \frac{k^{2}-q_{T}^{2}}{q_{T}}+2fk\right]\alpha
+\left[-2q_{L}+f\frac{k^{2}-q_{T}^{2}}{k}\right]\beta  =
\frac {k^{2}-q_{T}^{2}}{q_{T}} -2fk
\end{equation}
which yield  for the $T\rightarrow T$ and $T\rightarrow L$
 reflection coefficients :

 \begin{equation}
 \label {eq:rigCb3}
 \alpha = 1 + \frac{4fq_{L}}{\Delta}\hspace{100pt}\beta =
 \frac{4fk^{2}}{q_{T}\Delta}
 \end{equation}
 where :

 \begin {equation}
 \label{eq:rigCb4}
 \Delta = q_{L}\left(\frac{k}{q_{T}}+\frac{q_{T}}{k}\right) + f
 \left(q_{T} -2q_{L}-\frac{k^{2}}{q_{T}}\right)
 \end{equation}

 Note that, for $f\rightarrow 0+$, and for incident waves
 propagating towards $x_{1} > 0$ ($k > 0$),  $\Delta > 0$. That is, at least in
 the limit of weak friction and in this incidence range,
 clearly $\alpha >1, \beta > 0$, and the
 reflected energy flux is larger than the incident one.
 In order to try and clear up this a priori
 surprising prediction, let us look in more detail into the question of
 energy balance in our system. Consider the volume ($V$) of a slice of
 unit depth along $x_{3}$ of medium (M),
 limited by its area ($S$) ($-L<x_{1}<L$) along the interface, and the
 semicylinder ($C_{\infty}$) of radius $L (L\rightarrow\infty$).
In the situation we are considering, of an incident wave of constant amplitude,
the average over an acoustic period of the elastic energy stored
within ($V$) is time independent. This means that energy
conservation simply imposes that the net energy flux at
infinity ${\dot{W}}_{\infty}$, flowing out of $C_{\infty}$, associated with
the acoustic waves,
must equal the increment associated with the acoustic perturbation of
the work injected per unit time into the system via the
work of the stresses acting on the moving interface, ${\dot {W}}_{ext}$,
which is pumped from the driving machine. This is proved in
detail in the Appendix, where we show that, in the present case on a non
deformable M', per unit area of ($S$):

\begin{equation}
\label{eq:rigCb5}
\langle {\dot{W}}_{ext}\rangle  = \langle -{\dot u}_{1}
\tau_{12}\rangle
\end{equation}
where $\langle\ldots\rangle$ stands for the average over the acoustic
period. We then easily prove (Appendix A) that this is exactly equal
to the net acoustic energy flux :

\begin{equation}
\label{eq:rigCb6}
\dot{W}_\infty=\langle {\dot{ W}}_{refl}\rangle -\langle{\dot W}_{inc}\rangle 
\end{equation}

In other words, a sliding frictional interface is in principle able to
transform
external incoherent mechanical work into coherent acoustic radiation -
i.e. to act as an "acoustic laser". In order to make this more
precise, we plot on Figure \ref{fig:2}a the relative powers reflected into the T
and L channels, given by (see Appendix):

\begin {equation}
\label{eq:rigCb7}
w_{TT} = \mid\alpha\mid^{2}
\end{equation}

\begin{eqnarray}
\label{eq:rigCb8}
w_{TL} =\frac{q_{T}q_{L}}{k^{2}}\mid\beta\mid^{2}\hspace{50pt}if~ q_{L}~
real\\
=0\hspace{50pt}if ~q_{L}~ imaginary\nonumber
\end{eqnarray}
against the incidence angle $\theta =
\sin^{-1}\left(c_{T}k/\omega\right)$ and for $f = 0.2$, $c_{T}/c_{L} =
0.5$. It is seen that, for all positive incidences, a
transverse incident wave is predicted to be notably amplified upon
reflection, while
the power in the L channel remains quite small, even close to
$\theta_{lim} = \sin^{-1}(c_{T}/c_{L})$ beyond which the L wave
becomes evanescent, where it is maximum.
However a much more surprising result, already derived by Nosonovsky
and Adams \cite {Noso2} is that $w_{TT}$ and $w_{TL}$ are found to
diverge at a negative incidence $-\theta_{cr}$, where the
denominator $\Delta$ (Eq.(\ref{eq:rigCb4})) of the expressions for
$\alpha$ and $\beta$ (Eq.(\ref {eq:rigCb3}) vanishes. Using
equation (\ref{eq:qLT}), the condition $\Delta = 0$ reads :

\begin{equation}
\label{eq:rigCb9}
\sqrt{\frac{c_{T}^{2}}{c_{L}^{2}}-\sin^{2}{\theta}} =
-\frac{f\cos~2\theta \sin~\theta}{1-f\sin~2\theta}
\end{equation}
which is easily checked graphically to have a single negative
solution $ - \theta_{cr}$ with $\theta_{cr} < \theta_{lim}$, whatever
the values of $f$ and of $c_{T}/c_{L}$. The smaller $f$, the
closer $\theta_{cr}$ approaches $\theta_{lim}$. In the case shown on Figure
\ref{fig:2}a, $\theta_{lim}-\theta_{cr} \simeq 7'$.
In the ccase (Figure \ref{fig:2}b) of L incidence, the corresponding 
singularity occurs for $\theta$ close above $-90^{\circ}$
In other words, in this admittedly oversimplified limit
(rigid M', pure Coulomb friction), homogeneous sliding at constant
velocity is impossible : indeed, any infinitesimal perturbation is
able to trigger the emission, all along the interface, of a coupled
set of transverse and longitudinal acoustic waves, of a priori undetermined
amplitude, travelling
respectively at angles $\theta_{cr}$ and
$\sin^{-1}\left(\sqrt{\frac{c_{T}^{2}}{c_{L}^{2}\sin^{2}\theta_{cr}}-1}
\right)$ 
in
the back direction $-\hat{x}_{1}$. On the one hand this entails that
an infinite external
energy would be pumped in. On the other hand, as soon as the local
interfacial velocity will vanish, the interface will stop (stick). So
this pathologic behavior signals that homogeneous sliding is here
absolutely unstable - a signature to be added to that provided by the
existence in such systems of amplified interfacial waves, first
identified and studied by Adams \cite{Adams}, which lead to the question of
ill-posedness of the problem of interfacial slip pulses studied
by several authors and recently synthetized by Ranjith and Rice
\cite{Ranjith}.
Whether or not the stable mode of sliding for our model interface
would be a set of square slip pulses such as calculated in
ref.\cite{Noso2} is out of the reach of this work.
We will rather concentrate on a
different question, namely : how robust is this pathology? Does it
persist in the presence of (i) a finite elastic compliance of
the stiffer medium M'; (ii) a velocity-strengthening dependence of the
friction coefficient? We now consider successively these two
possibilities.

\subsection{ Finite elastic contrast and Coulomb friction}

Let us now consider the case of a large but finite elastic contrast
ratio $R = \mu'/\mu >>1$. We assume for the sake of simplicity
equality of the Poisson ratios $\nu = \nu'$.
In this case, except for quasi-normal incidence angles,
${q'}_{T,L}^{2} < 0$ : there
is total reflection at the interface, with only evanescent
transmission into M'. $\alpha$ and $\beta$ must now be obtained by
solving the full equations (\ref{eq:A1}) - (\ref{eq:A4}) for constant
$f$. Except in the vicinity of $\theta = -\theta_{cr}$, a
finite $R^{-1}$ simply acts as a regular perturbation, leading to a
correction ${\cal{ O}}\left(R^{-1}\right)$. For $\theta \approx
-\theta_{cr}$, however, one must consider in detail the behavior of
the determinant $\cal D$ of the system (\ref{eq:A1}) - (\ref{eq:A4}).
To first order in $R^{-1}$ :

\begin{equation}
\label{eq:rigCb10}
{\cal{D}} \propto  \Delta \left(q_{T}, q_{L}\right) + R^{-1} \Gamma
\left(q_{T}, q_{L}, {q'}_{T},{q'}_{L}\right)
\end{equation}
As seen above, the zero of $\Delta$ always occurs for $\theta_{cr} <
\theta_{lim}$ where $q_{L}$, and hence $\Delta$, is real. On the other
hand it appears that, for ${q'}_{T,L}$ pure imaginary, $\Gamma$ is a
complex quantity. So, no zero of ${\cal D}$  which would evolve
continuously from $-\theta_{cr}$ exists, and one finds that, for
finite $R^{-1}$, $\alpha$ and $\beta$ no longer diverge, but exhibit,
in the vicinity of $-\theta_{cr}$, a maximum of order $R$ : a finite
compliance of M', however small it is, is sufficient to kill the pathology
found in the totally rigid limit.
This we have confirmed by computing the relative reflected powers
$w_{IJ}\ (I,\,J=T,\,L)$
for various values of $R$. The results for $w_{TT}$ are shown on 
Figure \ref{fig:3}: the
smaller $R$, the lower the maximum of $w$, which reduces to a few
units for $R \lesssim 40$. Of course, in practice, for $R>>1$, the
very large values of the reflection coefficients at maximum mean that
a very small incident amplitude $\epsilon$ would suffice to induce a
finite response, such that $v+{\dot u}_{1}$ would vanish, leading to
interfacial stick - i.e. homogeneous slip, though no longer
absolutely unstable from a strict mathematical point of view, would be
very weakly stable, as it
could not resist
perturbations of finite but very small amplitude.

\subsection{Infinite elastic contrast and $v$-strengthening friction}

The problem is now specified by eqs. (\ref{eq:A1}) and (\ref{eq:A3})
in which $\alpha' = \beta' = 0$, and the position of the singularity
of $\alpha, \beta$, if any, is therefore given by the zero of:

\begin{equation}
\label{eq:rigCb11}
\Delta_{A} =
 q_{L}\left(\frac{k}{q_{T}}+\frac{q_{T}}{k}\right) + f
 \left(q_{T} -2q_{L}-\frac{k^{2}}{q_{T}}\right) +
 A\frac{\omega f(v)}{c_{T}}\left(\frac{q_{L}}{k}
 +\frac{k}{q_{T}}\right)
 \end{equation}
 where the dimensionless parameter measuring the velocity dependence
 effect is defined by equation (\ref{eq:A5}):

$$A = \frac{f'(v)c_{T}}{f(v)}\frac{\mid \tau^{*}_{22}\mid }{\mu}$$

When solving numerically for $\Delta_{A} = 0$, we find that the
singularity disappears for $ A \geq A_{cr}$. More precisely, we find
that, as $A$ increases, the zero of $\Delta_{A}$ approaches
$-\theta_{lim}$, which it reaches for $A = A_{cr}$, beyond which it
disappears, due to the fact that
no zero can occur in the regime where the L wave is evanescent.
The threshold
value $A_{cr}$ is independent of the value of $f(v)$, but it depends upon 
the sound velocity ratio. For a reasonable choice of
this parameter ($c_{T}/c_{L}< 0.7$), $A_{cr}$ is at
most a few units. For example, for the presently used value 
$c_{T}/c_{L} = 0.5$,
we obtain $A_{cr} = 1$. This behavior is illustrated on Figure \ref{fig:4},
where we
plot the relative reflected power $w_{TT}$ against incidence angle
for various values of $A$. It is also seen that the larger $A$, the
smaller the gain at reflection. It appears that at another characteristic 
$A$ value ($A=2$ for our parameter values), the gain becomes negative 
($w_{TT} < 1$) for all incidences.
That is, a strenghtening
$v$-dependence of $f$ is very efficient to kill the amplification
effect.

It is therefore important to evaluate an order of magnitude of $A$ for
real interfaces with a large elastic contrast, a good example of which
is that of gel/glass couples.
Frictional sliding at the interface between glass
and a 5\% gelatin aqueous gel has recently been studied in detail by
Baumberger et al \cite{Ronsin}. Using their data, we find that, for $v$
in the mm.sec$^{-1}$ range, $f'(v)/f(v) \approx 2.10^{3}$ sec.m$^{-1}$,
while $\left(c_{T}\mid{\tau}^{*}_{22}\mid/\mu\right) \approx 2$
m.sec$^{-1}$, so that  $A \approx 4.10^{3}$ is a very large number.

In this $A >> 1$ limit, the rigid M' version of equation
(\ref{eq:A3}) reduces, to lowest order in $A^{-1}$, to :

\begin{equation}
\label{eq:rigCb13}
\alpha + \beta + 1 = 0
\end{equation}
that is, due to the high relative cost of increasing its
instantaneous velocity, the interface remains locked to the
homogeneous sliding state and $u_{1} \cong 0$.

So, for gel/glass sytems,
acoustic reflection should not in practice be able to distinguish between the
static and the sliding interfaces.

Another realization of the high elastic contrast situation is
provided by the multicontact interface between rough glassy PMMA and
atomically flat silanized glass, recently studied by Bureau et al
\cite{Bureau}. For this
couple, $\mu'/\mu \simeq 20$. Thanks to the flatness of the glass
surface, it is possible to saturate the slow growth of the real area
of contact, which is responsible, in the case of rough-on-rough
systems, for the velocity weakening behavior of the dynamic friction
coefficient and of the associated stick-slip dynamics. Then,
for $v \gtrsim 1 \mu$m.sec$^{-1}$, $f(v)$ is velocity strenthening and
of the form :

\begin{equation}
\label{eq:rigCb14}
f(v) = f_{0}\left[1 + \zeta \log\left (\frac{v}{v_{0}}\right)\right]
\end{equation}
where   $f_{0} \approx 0.2$, and $\zeta \approx 2 - 4.10^{-2}$. With
$\mu_{PMMA} = 1$ GPa, $c_{T} \simeq 10^{3}\,$m.sec$^{-1}$, and under normal
stresses on the order of $5$ kPa, one gets :

\begin{equation}
\label{eq:rigCb15}
A \approx \frac{100 - 200}{v_{\mu m.sec^{-1}}}
\end{equation}
That is, for sliding velocities in the $100\, \mu$m.sec$^{-1}$ range,
$A$ is typically on the order of $1 - 2$.

Finally, note that, since in this configuration the average distance
between the
micrometric regions which form the real area of contact is on the
order of a fraction of millimeter,
such an experiment would ask
for acoustic signals in the range of of a few hundred kHz at most, in
order for our continuum description to be valid.

The relative powers reflected into the two channels for T and L
incident waves under these conditions are plotted on Figures \ref{fig:5} and 
\ref{fig:6},
for $A = 0, 1, 2$. (Since in this case typical values of
$\tau^{*}/\mu$ and $v/c_{T}$ are $< 10^{-5}$, calculations
have been performed
for $\tau^{*} = 0$, $\omega' = \omega$.) Inspection of these results and
comparison with the case of the non moving interface (Figure \ref{fig:7})
lead us to the following conclusions.

\begin{itemize}
\item
The more favorable channels for observing gain at reflection are the
diagonal (LL and TT) ones.
\item
In both cases, a very narrow peak of $w$,
reminiscent of the singularity obtained in the limit of Coulomb
friction on a rigid substrate, is predicted for $\theta \simeq -
\theta_{lim}$. However, due to its narrowness, observing it
would be certainly be excessively demanding in terms of directional
accuracy.
\item
It thus appears more feasible to investigate one of the two following
configurations : LL at large negative incidence angles, in
the range of $- 45 ^{\circ}$, and (TT) at large positive incidence $ \theta
\approx 60^{\circ}$.
\item
Gain decreases very rapidly with increasing $A$, i.e. with decreasing
$v$ (see eq.(\ref{eq:rigCb15})) : while, for $A = 1$
it reaches up to $20\%$ in (LL) and $35\%$ in (TT), already for $A
=2$ it has practically collapsed to zero. So, according to equation
(\ref{eq:rigCb15}), the best situation
corresponds to the largest possible driving velocities, in
practice in the range of a fraction of mm/sec.
\end{itemize}

With these conditions in mind, it seems quite possible to obtain
substantial acoustic gain at
reflection on a sliding rough-on-flat multicontact interface.

\section{Symmetric case}

We now turn to the opposite limit where M and M' have the same elastic
properties. It is a priori relevant to the case of multicontact interfaces
between two pieces of the same material. However, the above mentioned
rough-on-flat configuration is inappropriate in this case, since the
asperities on the rough surface would give rise on the flat one to
unavoidable and poorly qualified plastic damage and wear effects. On
the other hand, as already mentioned, in the rough-on-rough
configuration where such problems are irrelevant, geometric aging
interrupted by motion results in a velocity-weakening behavior of $f(v)$
for velocities up to at least the mm.sec$^{-1}$ range, not very
easy to acess in a stationary sliding laboratory experiment.
So, such symmetric solid on solid interfaces turn out not to be
well adapted in practice to test our predictions.

The most interesting case, as already mentioned in Section I, is
that of an established shear band, in particular in a highly confined
granular medium. In such highly disordered systems, internal stress
inhomogeneities (the so-called stress-chain phenomenon) of course give
rise to scattering of acoustic waves. However, this is all the smaller
that the effective
acoustic wavelength in the medium is larger with respect to the
correlation length of stress fluctuations, known to be on the order of
a few grain diameters $D$ only. In this $kD << 1$ limit, as shown
by experiments on the propagation of acoustic pulses \cite{Jia}, where one can
separate out unambiguously the signal corresponding to the
propagation of a "coherent pulse", a  description in terms of an
effective acoustic medium is legitimate. Note that, as shear band
thickness is also expected to be, for low shearing rates, on the order of a
few $D$, $kD << 1$ also
ensures that we can safely neglect it when dealing with our
reflection-transmission problem.

Laboratory experiments on
the frictional behavior of layers of granular rock confined between
granite plates \cite{Tullis} \cite{Marone}, and of glass bead
assemblies \cite{Geminard} have shown
that sheared confined granular media obey standard friction laws -
namely, beyond a static threshold, $\tau_{12} = - f \tau_{22}$.
However, the
velocity dependence of the dynamic friction coefficient is not yet
very clearly established. The data on rock point towards weakening
at low velocities. However, they are strongly dependent upon the level
of humidity. This, together with the very small size of the grains
used in these experiments, strongly suggests that capillary
condensation around the intergrain Hertz contacts is responsible for a
slow strengthening of the medium, interrupted by sliding, which
should become negligible either at low humidity or for larger grains.
On the other hand, in the absence of such slow transients, since
granular systems are essentially athermal, one expects
logarithmic dynamic strengthening of the type found for multicontact
interfaces (see equation (\ref{eq:rigCb15})), which results from
thermal activation effects \cite{Bo} \cite{Pauline2}, to be completely
negligible. If such is
the case, $f$ would be $v$- independent in the low velocity, quasi
static sliding regime. Note that this is what has be found by
Geminard et al \cite{Geminard} for glass beads completely immersed in water
- though under relatively weak confinement.

So, let us first consider the simplified case of pure Coulomb
friction ($A = 0$).
We then solve the corresponding version of eqs.(\ref{eq:A1}) -
(\ref{eq:A4}) where we neglect $\tau^{*}/\mu$, since, for
the confined granular systems made of hard materials (such as glass or
steel) which we have in
mind, under ordinary experimental conditions \cite{Jia} \cite{Gilles}, this
does
not exceed, typically, $ 10^{-5}$ at most. Moreover, as we
are interested here in strongly subsonic sliding velocities
($v/c_{T} < 10^{-5}-10^{-6}$), $\,\omega'
\simeq \omega$.

The results for
the four relative reflected powers
are plotted on Figure \ref{fig:8} for the case
$f = 0.2$ : while, in the (TT) channel, for
negative $\theta$ there is loss at reflection, $w_{TT}$ is seen to be
sizeable everywhere in the range ($-\theta_{lim}; \theta_{lim}$) where
a propagating L reflected wave exists, as well as for quasi-grazing
conditions. For $\theta = \pm \theta_{lim}$ the
reflectance curve exhibits cusp-like maxima - a standard behavior for
multichannel scattering cross sections at the closing point of a
channel. About $\theta = \theta_{lim}$, we even predict a gain of more
than $50\%$, which becomes much larger for $f = 0.4$. In the other
three channels, $w$ is much smaller - except, for the LL one, in quasi-grazing
conditions which are hardly realizable in practice.

This behavior is to be contrasted with what is expected from non
localized homogeneous shear sliding, namely undisturbed propagation all through
the sample. So, one expects the following to happen after starting to shear a
confined sample at a constant rate, if a transverse acoustic pulse is
sent into the granular pack in the direction normal to the shearing
plane. At very short times, when the stress has not
yet reached the sliding threshold, the system experiences uniform
elastic deformation, the incident pulse will give rise to a
"coherent" reflected one, which will have travelled way and back {\it
across the
whole container}. After the initial sliding transient corresponding to
the gradual installation of the shear band, during which
fluctuations are expected to kill the coherent signal, {\it the reflected
pulse should reappear, but with a distinctly shorter travelling time},
thus providing a clear signature of shear localization. Note that, since its
travelling length is reduced, so will be the attenuation due to scattering
by stress fluctuations, making it more easy to detect than the former
signal. Finally, when sliding is stopped, the situation should revert
practically
to that before shearing, though the density contrast
$\delta\rho/\rho$, on the order of a few $\%$ at most, associated with
the shear band \cite{Desrues} still persists. Indeed, then, $w_{TT}
\approx (\delta\rho/\rho)^{2}(d\omega/c_{T})^{2}$ is negligibly small,
and pulse reflection only occurs, again, at the bottom end of the
container.

The question is then to check how robust the reflectance
characteristics predicted for $A = 0$ are
with respect to a possible, though probably small,
velocity strengthening dependence of the
friction coefficient. It is seen on Figure \ref{fig:9} that,
although, as expected, $w_{TT}$ decreases as $A$ grows, at normal
incidence it remains non negligible up to the sizeable value $A = 2$,
where it still is on the order of $25\%$.

\section{Conclusion}
In summary, from the above results, we conclude to the interest of
performing experimental studies of the reflection of acoustic pulses
on sliding solid interfaces and shear bands, the potential interest
of which is different in each case.

As discussed in section III.C, experiments with a rough-on-flat
multicontact interface open the possibility of obtaining acoustic
gain, through the conversion by the frictional system of incoherent
mechanical energy pumped from the driving system into coherent
acoustic vibrations. However,  such experiments are certainly
delicate to realize, since they ask for working at non normal
incidences, and, due to the effect of
velocity strengthening of dynamic friction coefficients, at not very
small velocities, in the range of typically a few hundred $\mu$m/sec.

On the other hand, acoustic pulse reflection appears as a promising
method for detecting shear localization
in a confined granular
medium. The best configuration - a transverse pulse at
normal incidence - is more easily realizable since it may be
implemented with a single transducer acting as both emitter and
receiver.
Moreover, the expected signature (a strongly reduced transit time
before return of the reflected pulse) should be easy to identify - its
presence confirming at the same time the expected quasi-independence
of the sliding stress on the sliding velocity.

We therefore strongly hope that this work will motivate such
experiments in the near future.

\acknowledgements
B.V. gratefully acknowledges the hospitality of Universit\'e Paris VII.

\appendix
\section{Energy balance at a frictional interface\label{energy_balance}}

We restrict ourselves to the case of a rigid substrate with an
infinite elastic contrast $R=\mu'/\mu\rightarrow\infty$ considered in
Sec.~\ref{basic_case}. This situation is
sketched in Fig. \ref{fig:10}, which concerns specifically the case of a
transverse wave incident below the critical angle $\theta_{lim}$, 
so that also the
longitudinal reflected wave is propagating. In our 2D geometry, the 
incident beam irradiates a
strip $S_0$  of the interface perpendicular to the sliding 
direction $x_1$ and extending
from $-L_0$ to $L_0$ along this direction. It has an area $S_0$ 
per unit length along $x_3$. The
reflected  beams stem from the irradiated area. The beams are wide enough
to make the fringe effects negligible and easy to exclude as additive
constants to the principal quantities proportional to $S_0$. The $S_0$
strip is now overlapped by a wider region $S$ stretching from $-L$ to $L$ such
that a semicylindrical dome $C_\infty$ raised over it encloses the 
regions of beam
overlap and interference (as well as the fields of attenuated waves, should
they arise). The bottom of the dome is infinitesimally above the interface,
so that the equations of motion (\ref{eq:Lame1}), (\ref{eq:Lame2}) are 
valid both inside and on the surface of the region $V$ enclosed by 
the dome and the bottom plane.  Then, the total elastic
energy $W$ inside the dome can be studied using the differential energy
conservation law \cite{Hudson}, \cite{Morse} (see also \cite{Comninou})
\begin{equation}
\dyv{\jev}+\dot{e}=0\qquad\mbox{i.e.,}\qquad\jmath_{\ell,\ell}
+\dot{e}=0,\label{eq:ecl}
\end{equation}
where $e$ is the local energy density, whose form will not be needed, and
the energy current density $\jev$ is given by
\begin{equation}
\jev=-\dot{\uvec}\stress\qquad\mbox{i.e.,}\qquad 
\jmath_{\ell}=-\dot{u}_j{\tau}_{j\ell} 
\label{eq:defj}
\end{equation} 
We consider elastic fields having a steady homogeneous component and a
single frequency acoustic component. As usual, in bilinear expressions like
Eq. (\ref{eq:defj}), we turn to real parts of all quantities involved,
which is consistent in a linear theory. Thus, we have
\begin{equation}
\dot{\uvec}(\rvec,t)=\Re(-i\omega\uvec^{0}(\rvec){\rm e}^{-i\omega t})
\label{eq:real_u}
\end{equation}   
\begin{equation}
{\stress}(\rvec,t)=\stress{}^{*}+\Re(\delta\stress{}^{0}(\rvec)
{\rm e}^{-i\omega t})
\label{eq:real_tau}
\end{equation}   
Introducing this into (\ref{eq:defj}), we may single out the steady flow
component of $\jev$ by averaging over the period $2\pi/\omega$:
\begin{equation}
\jav=\half(-i\omega\uvec^{0}\overline{\delta\stress^{0}}
+i\omega\overline{\uvec^{0}}\delta\stress^{0})
\label{eq:jav}
\end{equation}    
where overbar denotes complex conjugation.
The time average $\langle W\rangle$  of the total elastic energy 
per unit length of the $V$ region for a stationary irradiation is zero.
Integration of the time average of Eq. (\ref{eq:ecl}) then yields a balance
equation for the steady time averaged energy flows across the surface of
$V$:
\begin{equation}
\int\limits_{C_\infty}dS\,\jav\cdott\nvec=
\int\limits_{S}dS\,(-\jav)\cdott\nvec
\label{eq:bal}
\end{equation} 
with $\nvec$ the outer normal of $V$. With the sign conventions of Figure
\ref{fig:10}, we obtain the following relation for the gain
$\dot{W}_\infty$, defined as 
the imbalance between the incident and reflected energy flows:
\begin{equation}
\dot{W}_\infty=\underbrace{\langle\dot{W}_T\rangle+\langle\dot{W}_L\rangle}_
{\displaystyle \langle\dot{W}_{refl}\rangle}
-\langle\dot{W}_{inc}\rangle=
\langle\dot{W}_{ext}\rangle
\label{eq:gaindef}
\end{equation} 
The energy flows associated with the plane wave beams crossing the dome
$C_\infty$ are easily obtained, if we use the well known steady current
density for a plane wave \cite{Hudson}
$$
|\jav_{L,T}|=\half\rho\omega^2c_{L,T}\cdot|U|^2
$$
valid for both polarizations; the quantity $U$ is a possibly complex
amplitude. The plane waves appearing in (\ref{eq:gaindef}) are defined
in Eqs. (\ref{eq:u1}), (\ref{eq:u2}). The coefficients $\alpha,\beta$ were
obtained in Sec.~\ref{basic_case}. The three total energy flows are 
the products of current densities and beam cross sections. Thus,
\renewcommand{\arraystretch}{1.8}
\begin{equation}
\begin{array}{rcl}
\langle\dot{W}_{inc}\rangle&=&\half\rho\omega^2c_{T}\cdot\epsilon^2
\displaystyle\left({\omega \over
{c_Tq_T}}\right)^2\cdot S_0{{c_Tq_T}\over\omega}\\
\langle\dot{W}_{T}\rangle&=&\half\rho\omega^2c_{T}\cdot\epsilon^2
|\alpha|^2\displaystyle\left({\omega \over
{c_Tq_T}}\right)^2\cdot S_0{{c_Tq_T}\over\omega}\\ 
\langle\dot{W}_{L}\rangle&=&\half\rho\omega^2c_{L}\cdot\epsilon^2
|\beta|^2\displaystyle\left({\omega \over
{c_Lk}}\right)^2\cdot S_0{{c_Lq_L}\over\omega} 
\end{array}
\label{eq:flows}
\end{equation}
\renewcommand{\arraystretch}{1.0}
The ratios of the flows $w_{TJ}=\langle\dot{W}_{J}\rangle/
\langle\dot{W}_{inc}\rangle$ $(J=T,L)$ then have the form given in 
Eqs. (\ref{eq:rigCb7}),(\ref{eq:rigCb8}) in the main text. 

It remains to interpret the right hand side $\langle\dot{W}_{ext}\rangle$ 
of (\ref{eq:gaindef}). First, using explicit expressions (\ref{eq:defj}),
(\ref{eq:jav}) for
$\jev$, $\jav$, we obtain (cf. (\ref{eq:rigCb5}))  
\begin{equation}
\langle\dot{W}_{ext}\rangle=\int\limits_{S_0}dS\,\langle\dot{u}_1\tau_{12}\rangle
\label{eq:wext1}
\end{equation}
\begin{equation}
\langle\dot{W}_{ext}\rangle=\int\limits_{S_0}dS\,
\half(-i\omega u_1^{(0)}\overline{\delta\tau^{(0)}_{12}}
+i\omega\overline{u_1^{(0)}}\delta\tau^{(0)}_{12})
\label{eq:wext2}
\end{equation}
Eq. (\ref{eq:wext2}) may serve to check equation (\ref{eq:gaindef}) 
explicitly.
Two points appear 
explicitly : {\it (i)} Only those irradiated parts of the interface where
$\dot{u}_1\neq 0$ do contribute: the effect is connected with the free
sliding of the interface points along the sliding direction. {\it (ii)} All
three waves superimposed enter $\langle\dot{W}_{ext}\rangle$. This is
particularly remarkable in the case of an evanescent L wave, in which 
there is no
longitudinal flow at infinity, yet the energy gain at the interface 
cannot be obtained correctly without including the L wave 
contribution right at the
interface.

At each point of the interface, the external force acting on medium $M$
is 
$$\Fvec=\stress\nvec=\stress^*\nvec+\Re(\delta\stress)\nvec
\equiv \Fvec^* + \delta \Fvec$$
The time average is
$$\langle\Fvec\rangle=\Fvec^*$$
The averaged macroscopic force acting on $S$ is thus ${\cal F}_S=S\Fvec^*$.
Medium $M'$ is pulled at the overall velocity $\vvec$, thus the power
spent by this force is ${\cal F}_S\cdott \vvec$, independently of the
presence of oscillatory acoustic fields. It is now easy to see that
\begin{equation}
{\cal F}_S\cdott \vvec=\int\limits_SdS\,
\langle(\vvec-\dot{\uvec})\cdott\Fvec\rangle +\langle\dot{W}_{ext}\rangle
\label{eq:power}
\end{equation}
The first term is the power dissipated against the friction forces:
$\vvec-\dot{\uvec}$ is the local relative interfacial velocity, 
while $\Fvec$ has only a frictional component along the interface.
Equation (\ref{eq:power}) thus expresses the partitioning of the total work
done per unit time by the external force into the (irreversibly) dissipated 
power and the net acoustic gain.

\newpage

%
\begin{figure}
\caption{Schematic representation of the system: a transverse incident wave
impinges at incidence $\theta$ onto the sliding interface, giving rise to
two reflected and two transmitted waves.
}
\label{fig:1}
\end{figure} 

\begin{figure}
\caption{Relative powers $w_{IJ}(\theta)\ (I,J=T,L)$ reflected from the
interface between a compliant medium and a rigid one versus incidence angle
$\theta$.  Friction is
velocity independent. $R=\infty,\,f=0.2,\,c_T/c_L=0.5$.
Incident wave  (a) transverse, $w_{TT}$ (---), $w_{TL}$ (-~-~-)  (b)longitudinal
, $w_{LL}$ (---), $w_{LT}$ (-~-~-).
}
\label{fig:2}
\end{figure}  

\begin{figure}
\caption{Evolution of the relative reflected power $w_{TT}(\theta)$ with
elastic contrast ratio $R$ for Coulombic friction. $f=0.2,\,c_T/c_L=0.5$,
$R= \infty$ ($\cdot\cdot\cdot\cdot$), 40 (-~-~-), 20 ($-~\cdot$~-~$\cdot$), 
10 (---).
(a) Full incidence range; (b) Blow-up for $\theta$ close to 
$-\theta_{lim}=-30^\circ$.
}
\label{fig:3}
\end{figure}

\begin{figure}
\caption{Evolution of the relative reflected power $w_{TT}(\theta)$ with 
strengh $A$ of the velocity depence of the friction coefficient 
for rigid $M'$. 
 $f=0.2,\,c_T/c_L=0.5,\,R= \infty$. $A=$ 0 (---), 1  (-~-~-), 
 2 ($-~\cdot$~-~$\cdot$).(a) Full incidence range; (b) Blow-up for $\theta$ close to
 $-\theta_{lim}=-30^\circ$. 
 Two more curves, for $A=0.5$  ($\cdot\cdot\cdot\cdot$), 0.9 (thin 
 full line) show the shift and narrowing of the singularity as $A=A_{cr}=1$
 is approached. For $A\ge A_{cr}$, the peak amplitude becomes
 finite. 
}
\label{fig:4}
\end{figure}  

\begin{figure}
\caption{Evolution of relative reflected powers with increasing $A$. Incident 
$T$~wave, 
$R=20,\,f=0.2,\,c_T/c_L=0.5$. $A=$ 0 (---), 1  (-~-~-),
 2 ($-~\cdot$~-~$\cdot$).  (a) $w_{TT}(\theta)$, (b) $w_{TL}(\theta)$.
 The height of the peak for $\theta=-\theta_{lim}$ is maximum, but 
 finite, for $A=1$.
}
\label{fig:5}
\end{figure}

\begin{figure}
\caption{Same plot as in Fig. \ref{fig:5}, but for an incident $L$ wave.
(a) $w_{LL}(\theta)$, (b) $w_{LT}(\theta)$.
}
\label{fig:6}
\end{figure}

\begin{figure}
\caption{Static interface: relative powers reflected from incident wave 
(a) transverse, $w_{TT}$ (---), $w_{TL}$ (-~-~-)  (b)longitudinal
, $w_{LL}$ (---), $w_{LT}$ (-~-~-).  $R=20,\,c_T/c_L=0.5$
}
\label{fig:7}
\end{figure}

\begin{figure}
\caption{ lastically symmetric case ($R=1$), and Coulombic 
friction ($A=0$): relative powers reflected from incident wave 
(a) transverse, $w_{TT}$ (---), $w_{TL}$ (-~-~-)  (b)longitudinal
, $w_{LL}$ (---), $w_{LT}$ (-~-~-).  
$f=0.2,\,c_T/c_L=0.5$. 
}
\label{fig:8}
\end{figure}

\begin{figure}
\caption{Evolution of the relative reflected power $w_{TT}(\theta)$ with
strengh $A$ of the velocity dependence of the friction coefficient 
for symmetric case $R=1$.
 $f=0.2,\,c_T/c_L=0.5$. $A=$ 0 (---), 1  (-~-~-),
  2 ($\cdot$~-~$\cdot$).
}
\label{fig:9}
\end{figure}

\begin{figure}
\caption{Energy flows for rigid $M'$.
}
\label{fig:10}
\end{figure} 

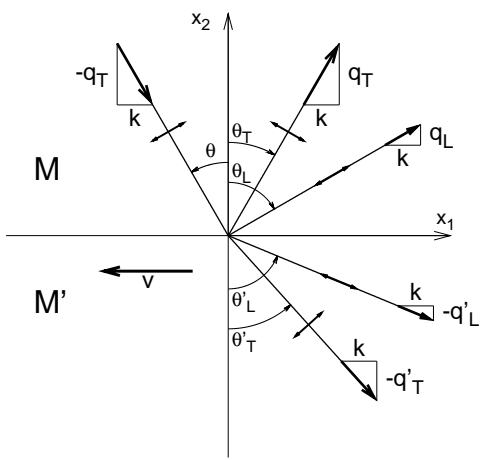
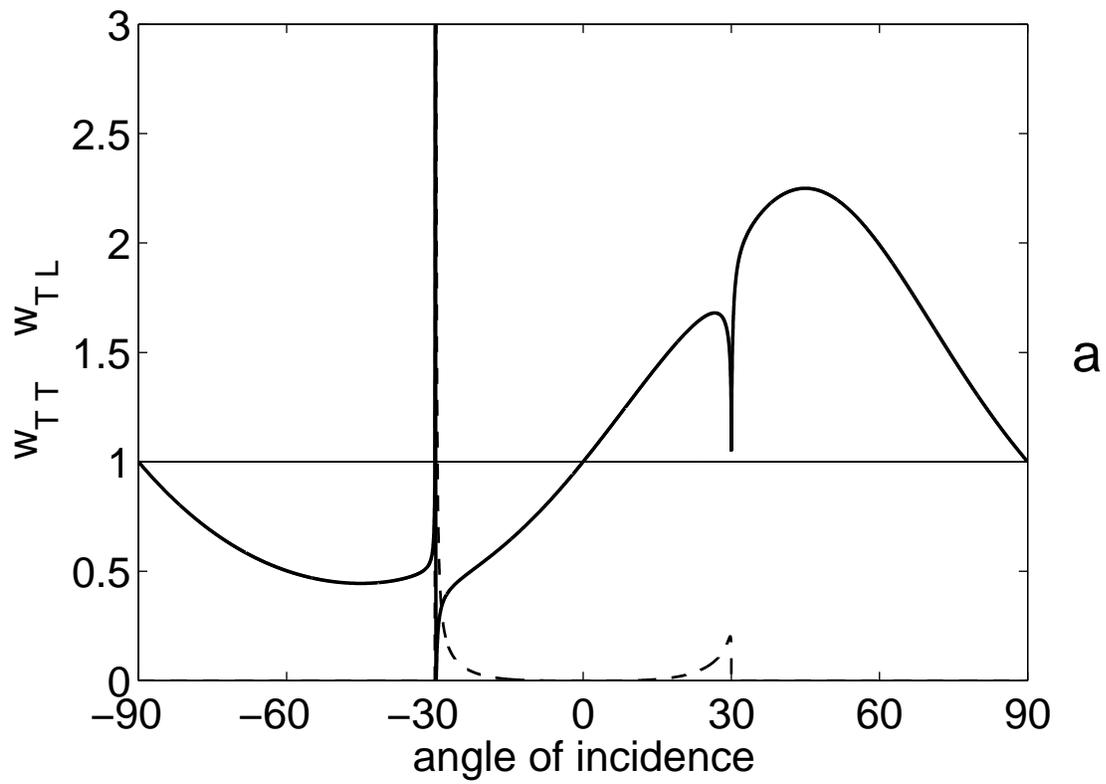
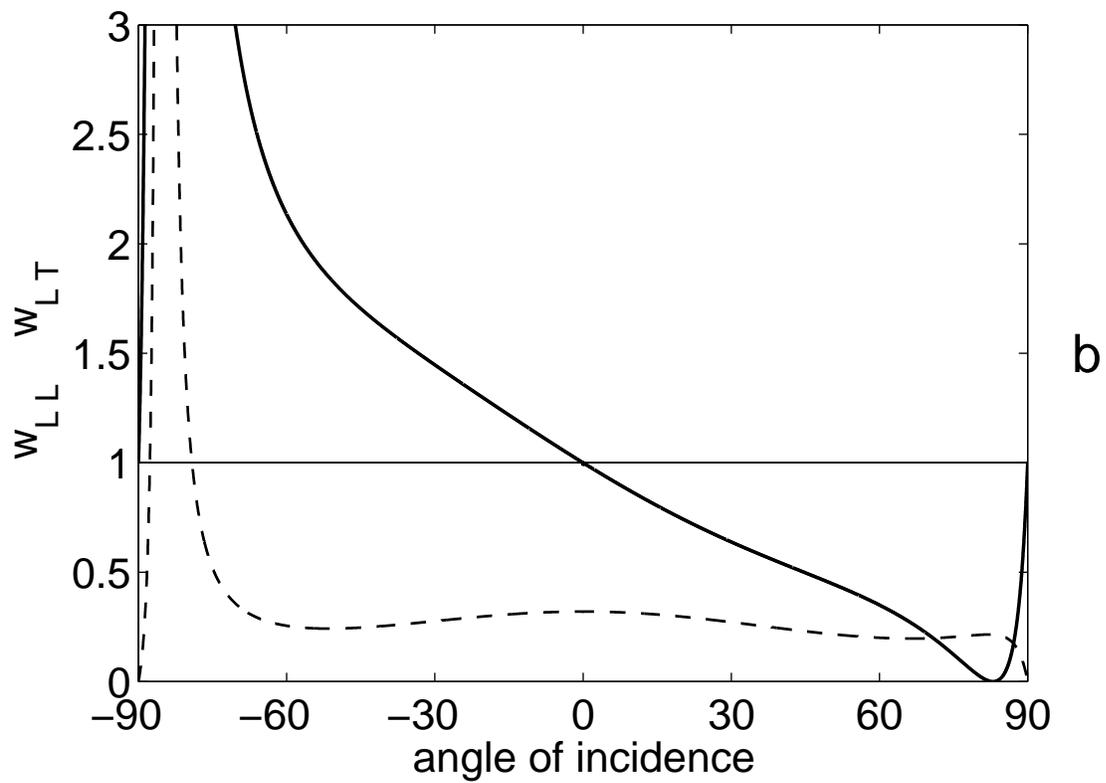
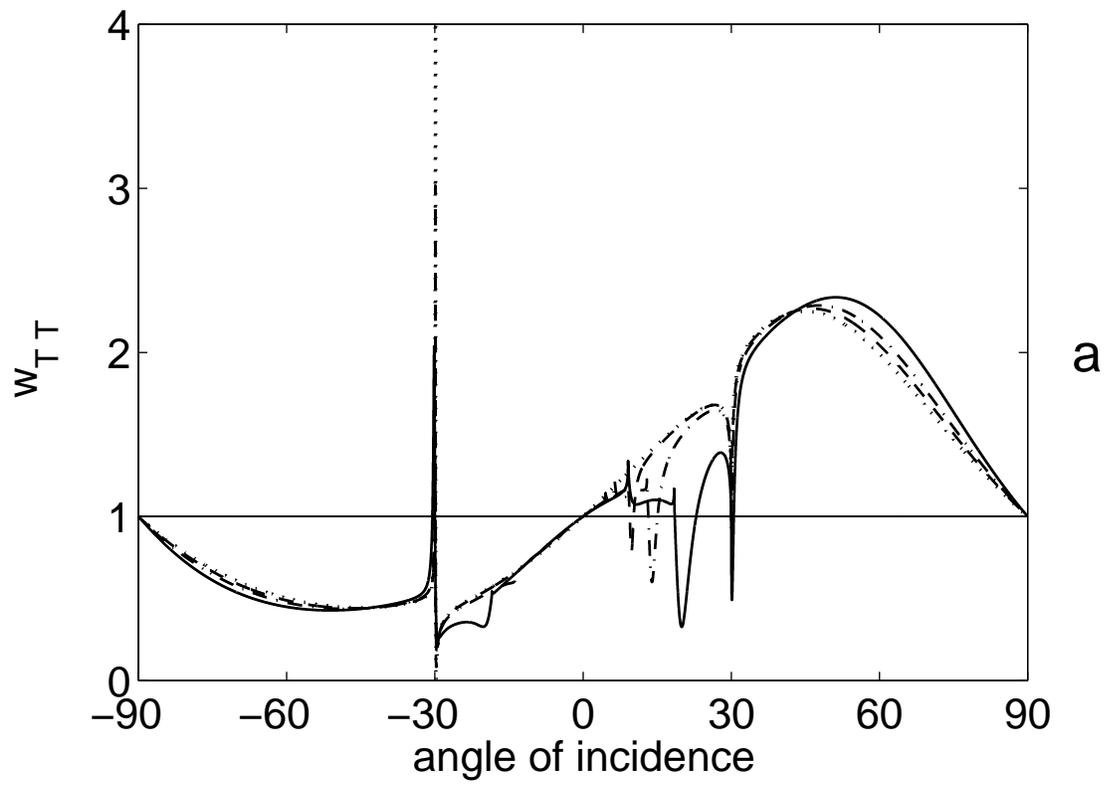
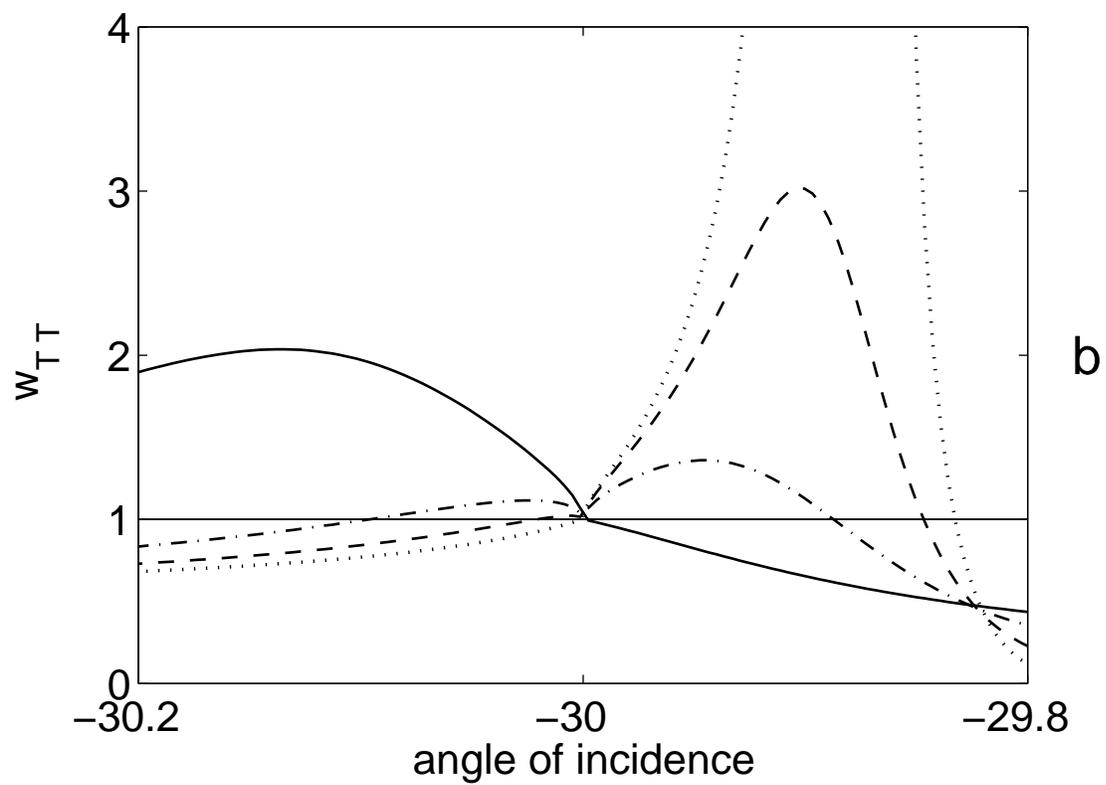
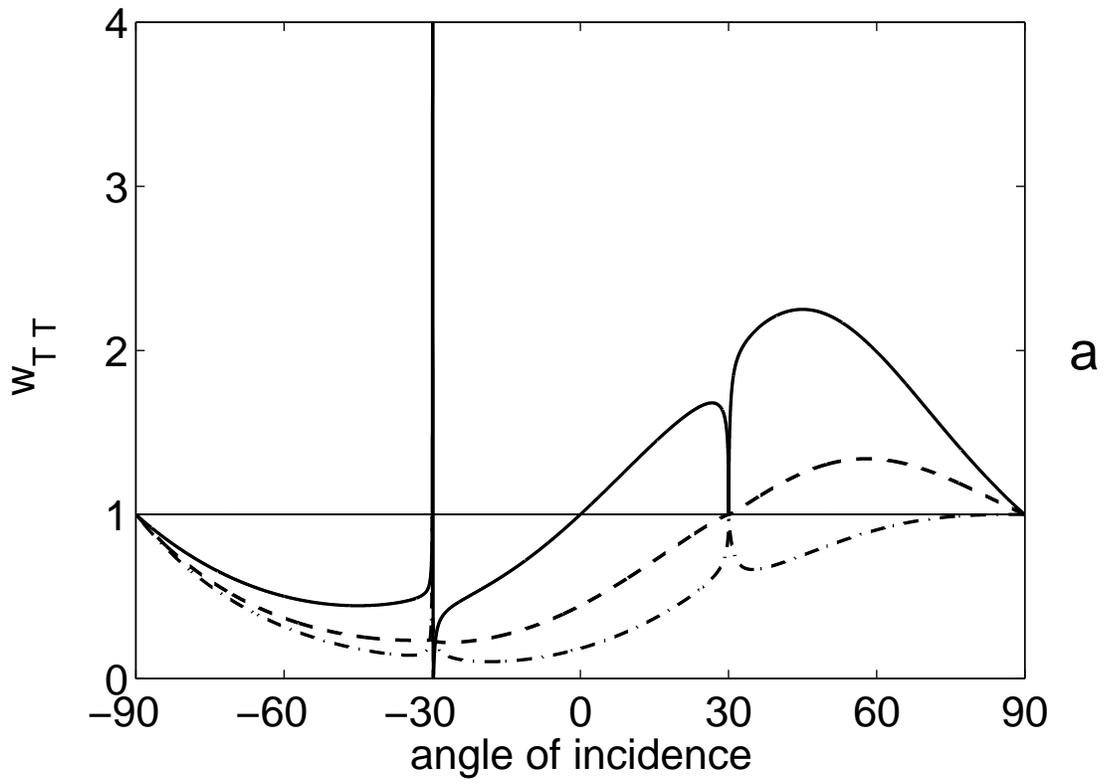
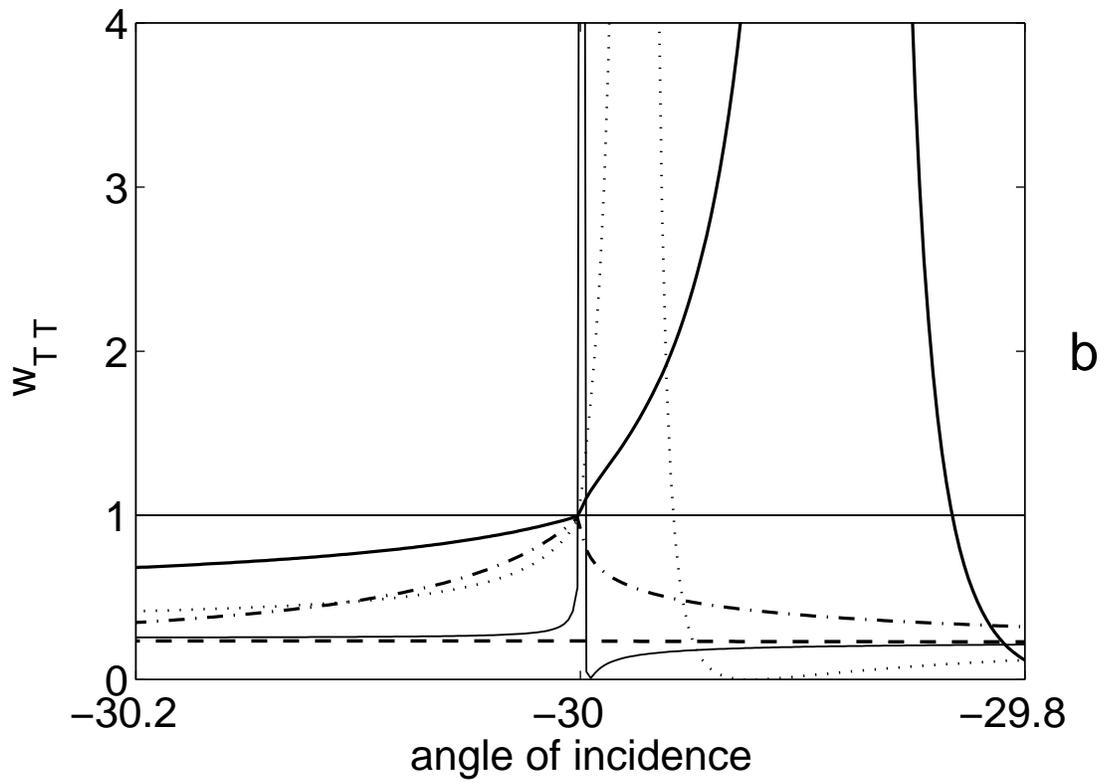
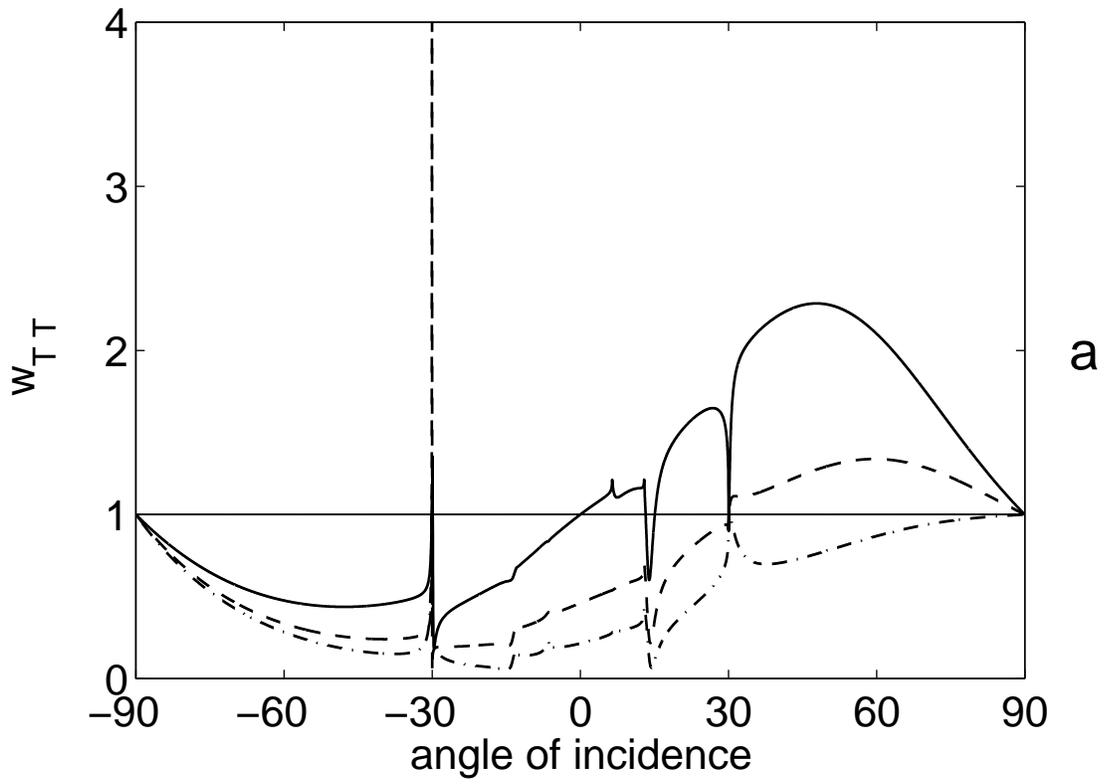
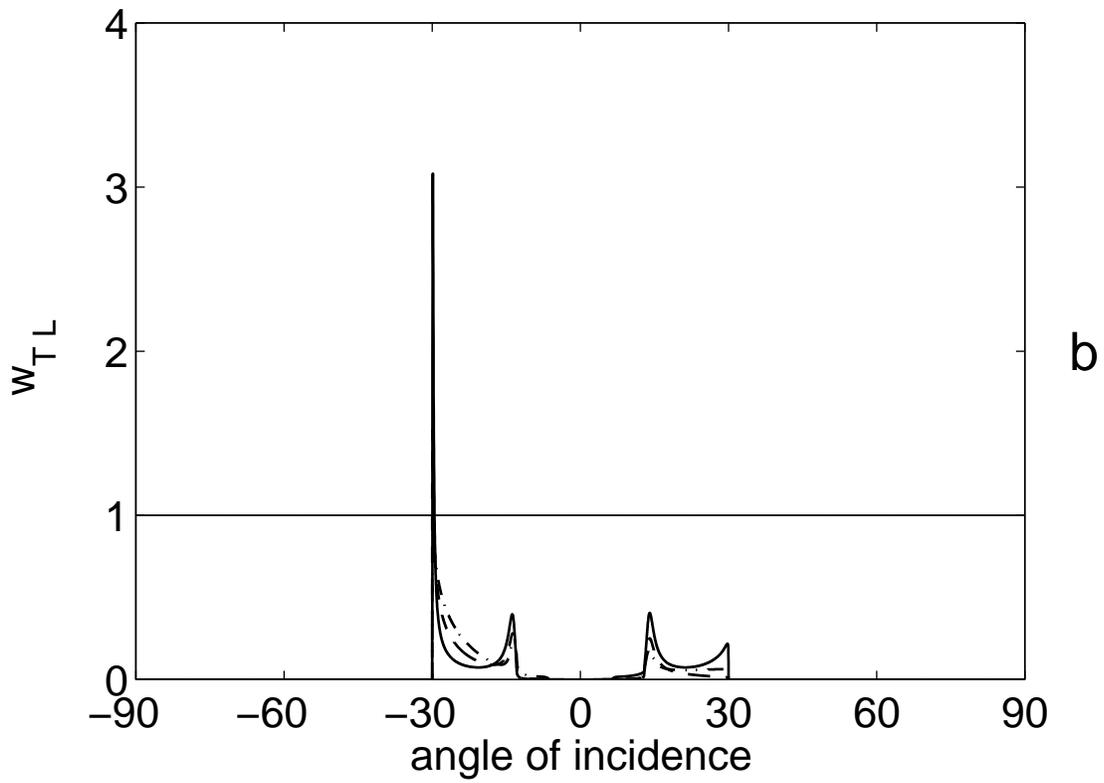
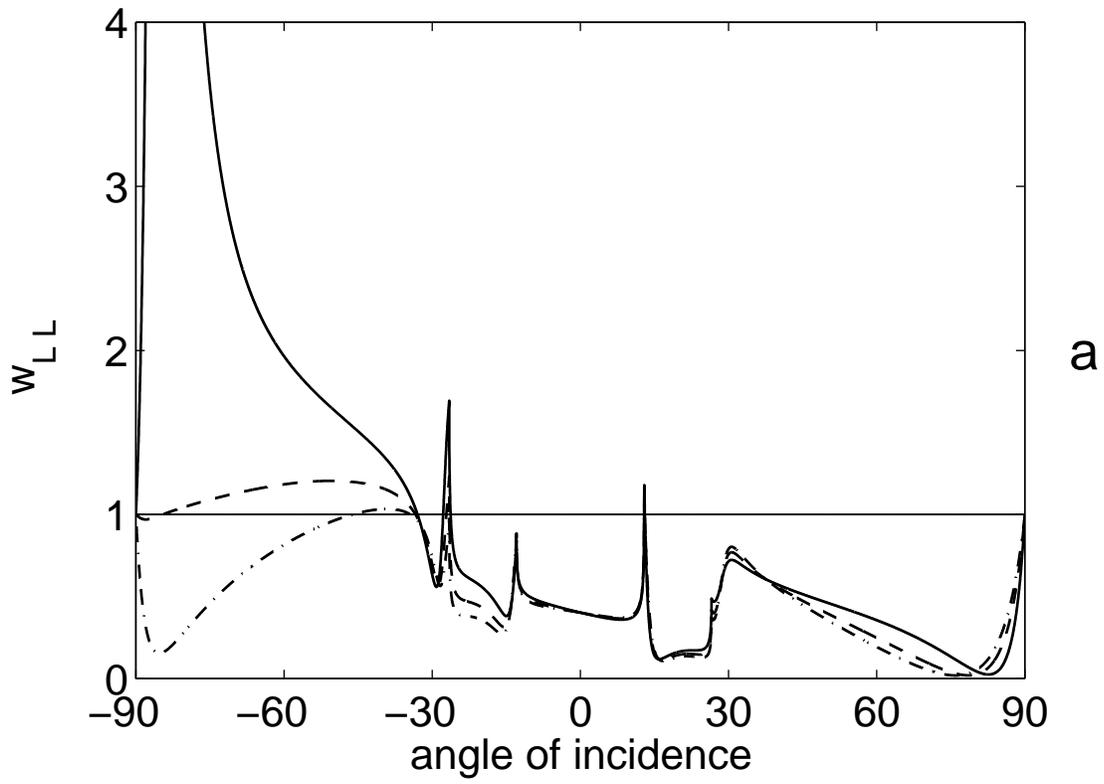
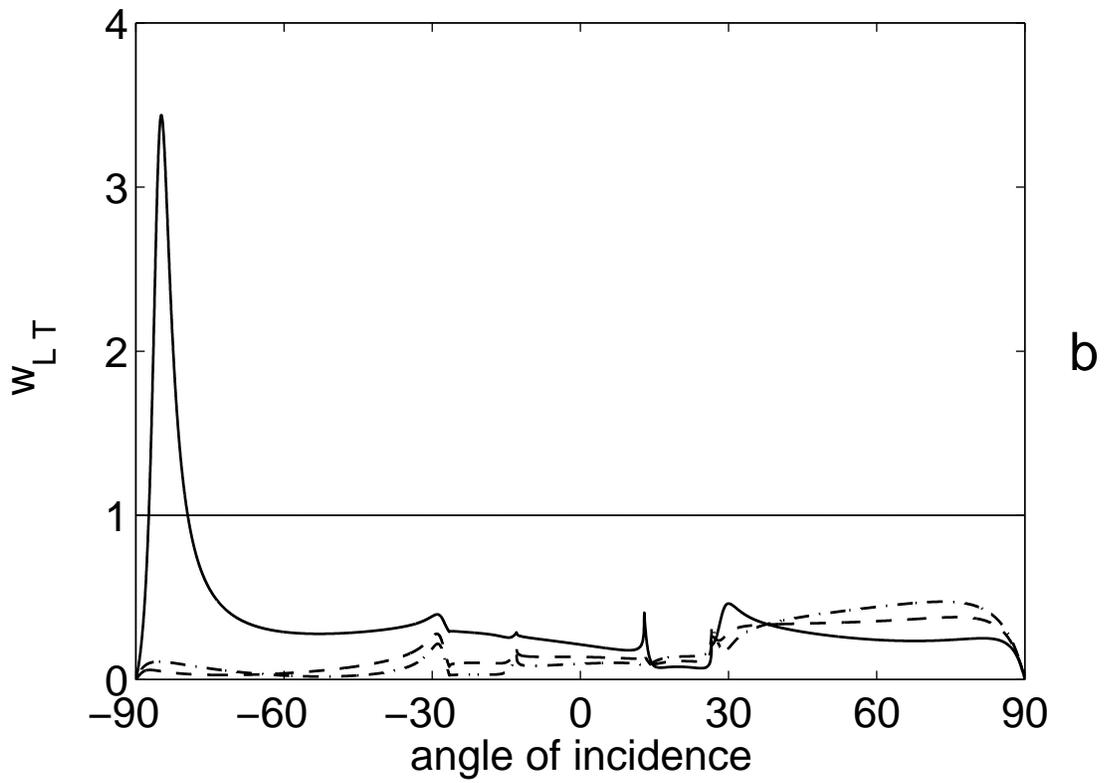
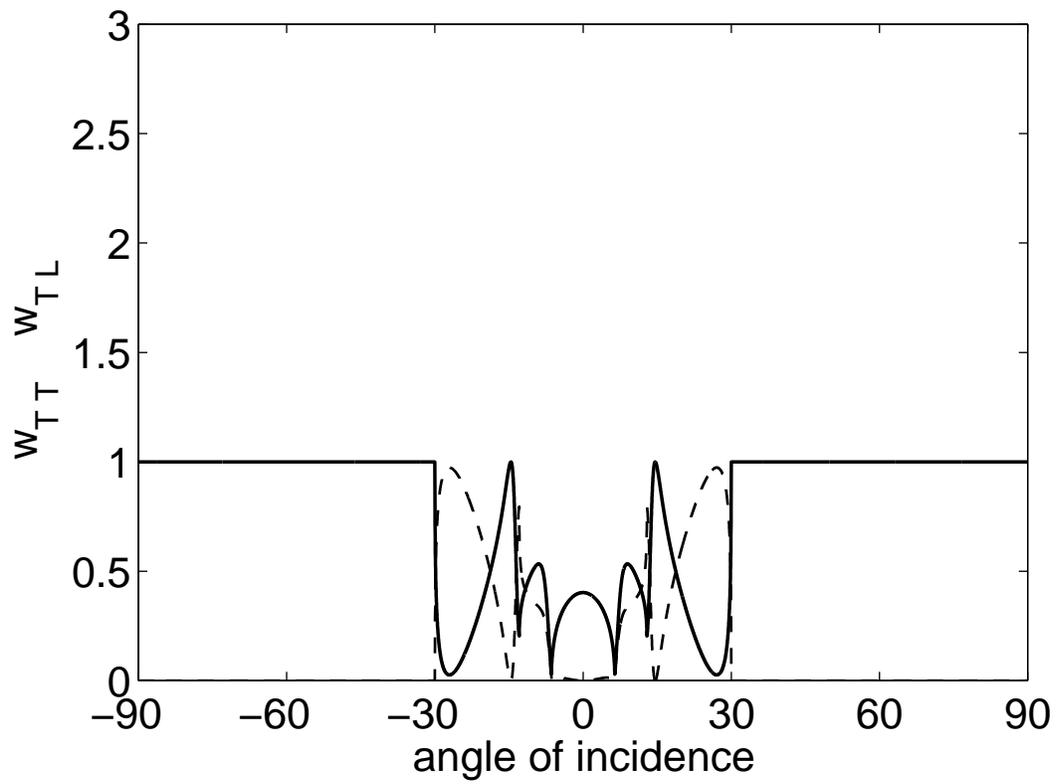
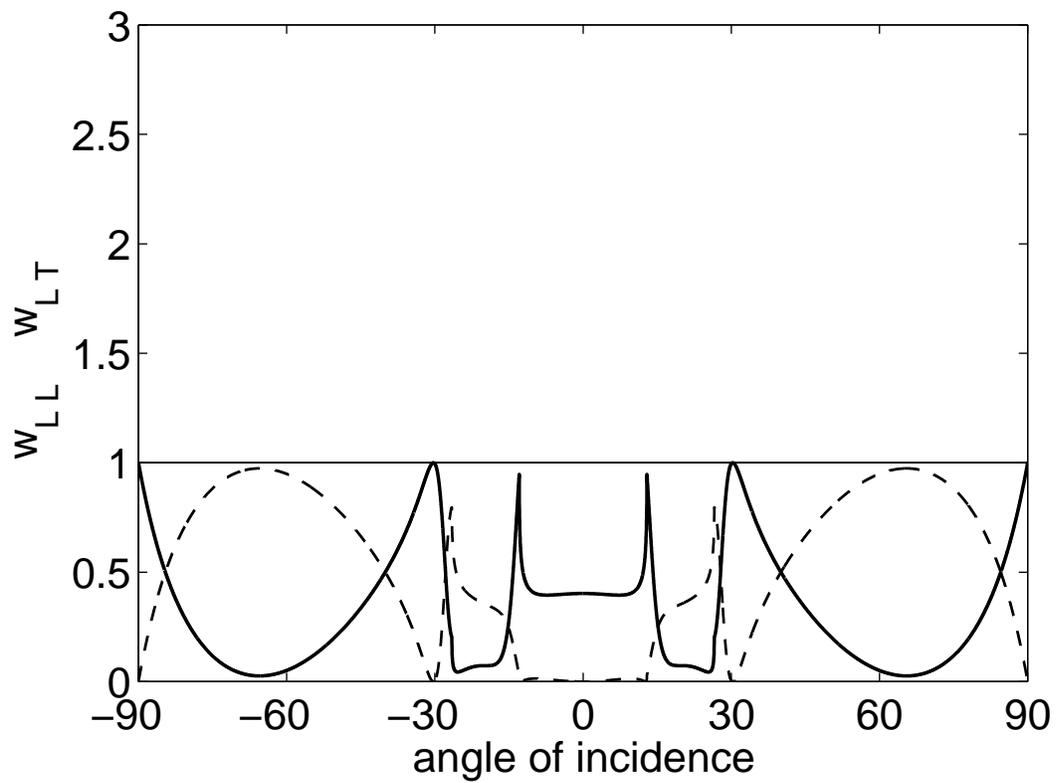
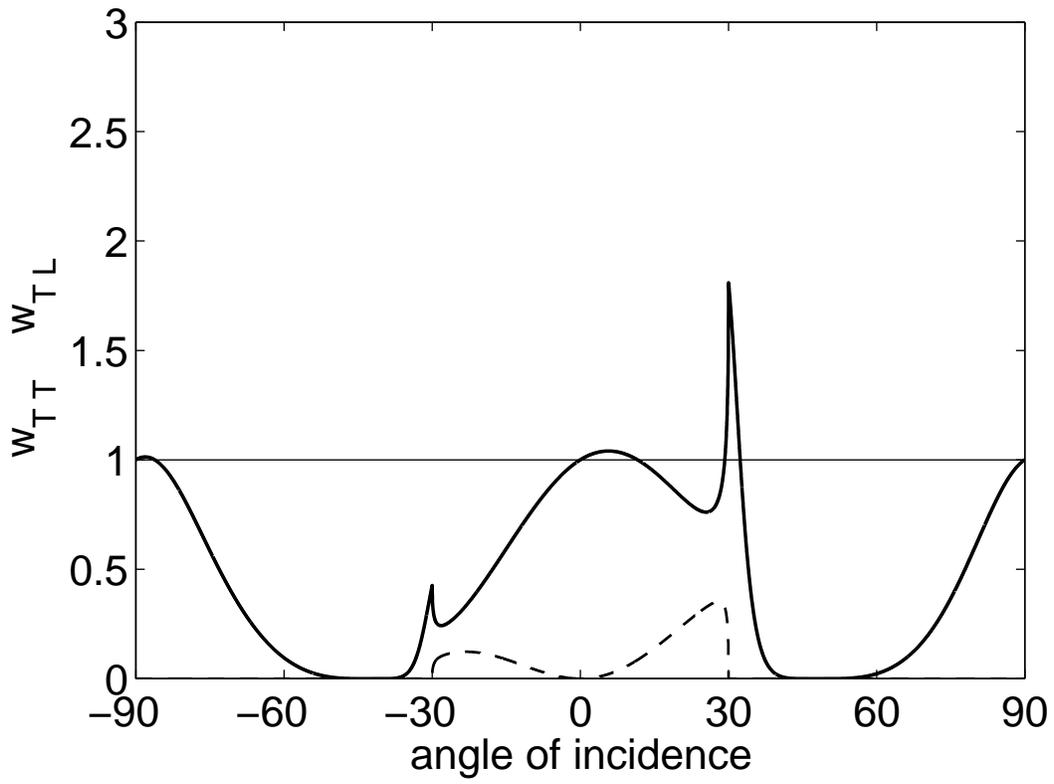
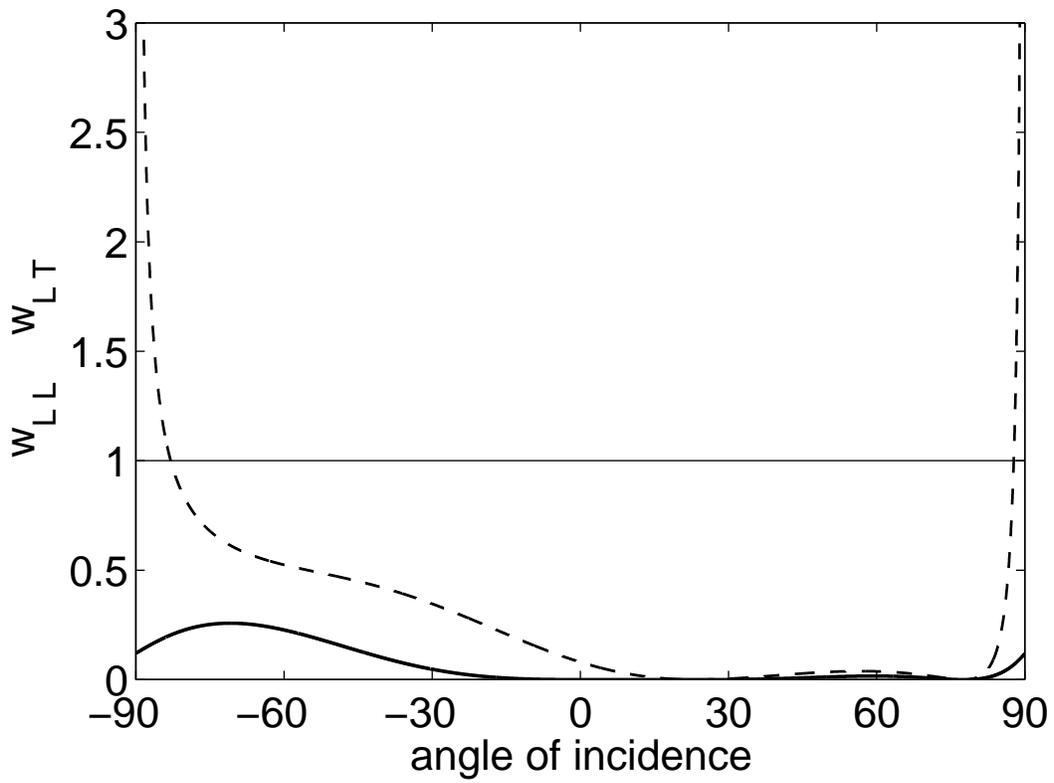
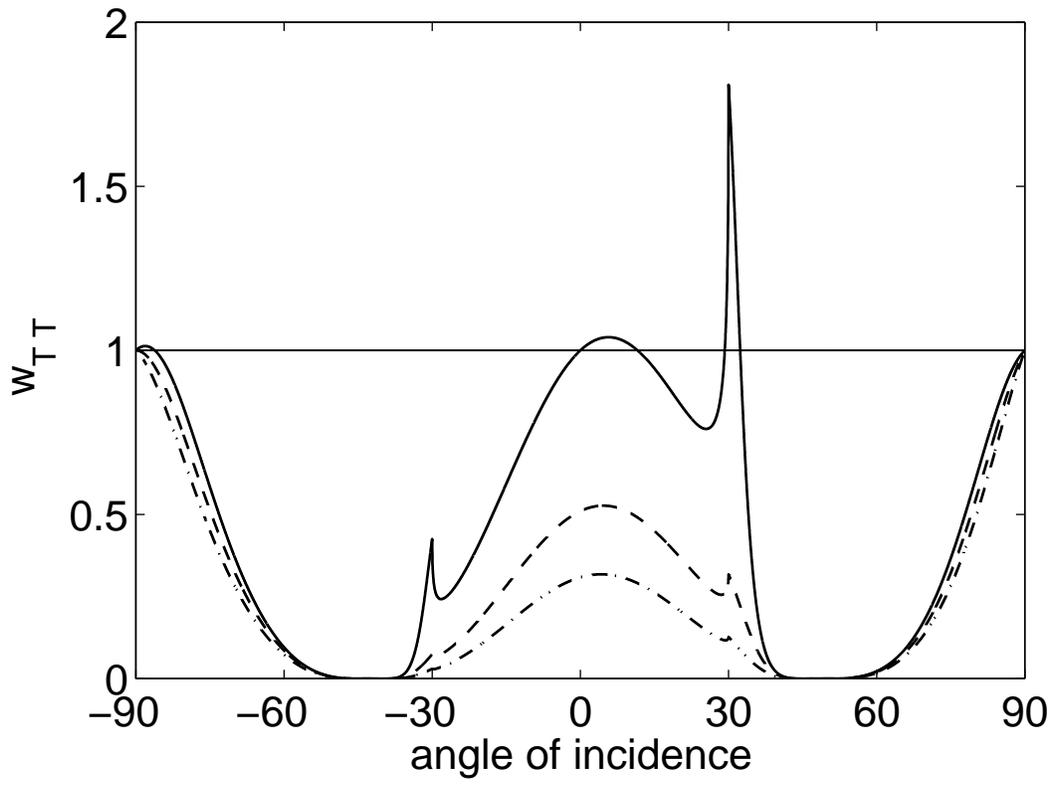
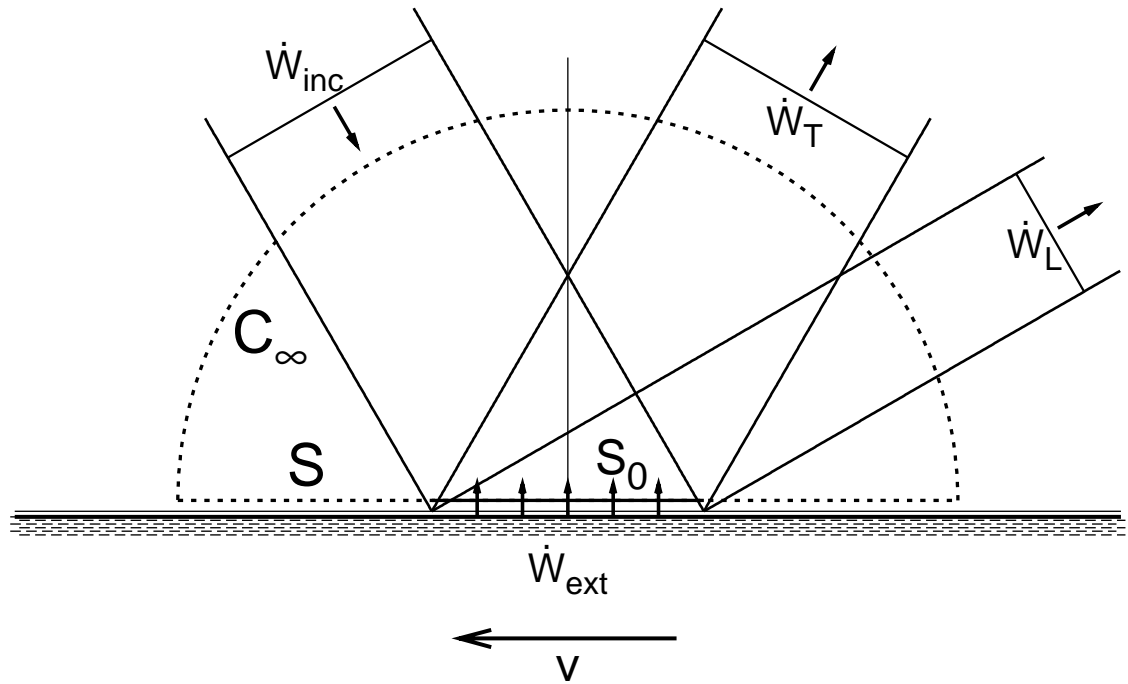
\end{document}